\documentstyle[12pt]{article}

\newlength{\dinwidth}
\newlength{\dinmargin}
\begin{document}

\def\bold#1{\setbox0=\hbox{$#1$}%
     \kern-.025em\copy0\kern-\wd0
     \kern.05em\copy0\kern-\wd0
     \kern-.025em\raise.0433em\box0 }
\def\slash#1{\setbox0=\hbox{$#1$}#1\hskip-\wd0\dimen0=5pt\advance
       \dimen0 by-\ht0\advance\dimen0 by\dp0\lower0.5\dimen0\hbox
         to\wd0{\hss\sl/\/\hss}}
\def\lq{\left [}
\def\rq{\right ]}
\def\LL{{\cal L}}
\def\VV{{\cal V}}
\def\AA{{\cal A}}
\def\MM{{\cal M}}

\newcommand{\be}{\begin{equation}}
\newcommand{\ee}{\end{equation}}
\newcommand{\bea}{\begin{eqnarray}}
\newcommand{\eea}{\end{eqnarray}}
\newcommand{\nn}{\nonumber}
\newcommand{\dd}{\displaystyle}
\newcommand{\bra}[1]{\left\langle #1 \right|}
\newcommand{\ket}[1]{\left| #1 \right\rangle}
\thispagestyle{empty}
\vspace*{1cm}
\rightline{BARI-TH/96-235}
\rightline{May 1996}
\vspace*{2cm}
\begin{center}
  \begin{Large}
  \begin{bf} 
On the  Decay Mode 
$B^- \to \mu^- {\bar \nu}_\mu \gamma$
\\  \end{bf}
  \end{Large}
  \vspace{8mm}
  \begin{large}
P. Colangelo $^{a,}$ \footnote{e-mail address:
COLANGELO@BARI.INFN.IT},  F. De Fazio $^{a,b}$, G. Nardulli $^{a,b}$ 
\footnote{e-mail address:
NARDULLI@BARI.INFN.IT}\\
  \end{large}
  \vspace{6mm}
$^{a}$ Istituto Nazionale di Fisica Nucleare, Sezione di Bari, Italy\\
  \vspace{2mm}
$^{b}$ Dipartimento di Fisica, Universit\'a 
di Bari, Italy \\

\end{center}
\begin{quotation}
\vspace*{1.5cm}
\begin{center}
  \begin{bf}
  ABSTRACT
  \end{bf}
\end{center}
\vspace*{0.5cm}
\noindent
A QCD relativistic potential model is employed to compute the decay rate 
and the photon spectrum of the 
process $B^- \to \mu^- {\bar \nu}_\mu \gamma$. The result
${\cal B}(B^- \to \mu^- {\bar \nu}_\mu \gamma) \simeq 1  \times 10^{-6}$ 
confirms the
enhancement of this decay channel with respect to the purely leptonic mode, and
supports the proposal of using this process to access relevant 
hadronic quantities such as the $B$-meson leptonic decay constant and the CKM 
matrix element $V_{ub}$.
\end{quotation}

\newpage
\baselineskip=18pt
\setcounter{page}{1}

Noticeable theoretical attention has been recently given to 
the weak radiative decay
\be
B^- \to \mu^- {\bar \nu}_\mu \gamma \;\;\;. \label{channel}
\ee
\noindent 
The reason is in the peculiar role of this decay mode for the
understanding of the dynamics of the annihilation
processes occuring in heavy mesons \cite{burdman,atwood,eilam,noi}.
Moreover, it has been observed 
that (\ref{channel}) can be studied to obtain indications
on the value of the $B$-meson leptonic 
constant $f_B$ using a decay channel which differs from the purely
leptonic modes $B^- \to \ell^- \;  {\bar \nu}_\ell$, 
and is not hampered by the limitations 
affecting those latter processes. Such difficulties mainly consist 
in low decay rates 
\footnote{Present bounds are:
${\cal B} (B^- \to e^- {\bar \nu}_e) < 1.5 \; 10^{-5}$, 
${\cal B} (B^- \to \mu^- {\bar \nu}_\mu) < 2.1 \; 10^{-5}$ \cite{cleo1} .}
(using $V_{ub}=3\times 10^{-3}$, $f_B=200 \; MeV$ and 
$\tau_{B^-}=1.646 \pm 0.063 \; ps$ \cite{hon} one predicts
${\cal B} (B^- \to e^- {\bar \nu}_e) \simeq 6.6 \; 10^{-12}$ and 
${\cal B} (B^- \to \mu^- {\bar \nu}_\mu) \simeq 2.8 \; 10^{-7}$ 
) 
or in reconstruction problems for
$B^- \to \tau^- {\bar \nu}_\tau$. 

In ref. \cite{noi} heavy quark symmetry and experimental data on $D^{*0} \to 
D^0 \gamma$ have been exploited to study the dependence of
 ${\cal B}(B^- \to 
\mu^- {\bar \nu}_\mu \gamma)$ on the heavy meson decay constant
 ${\hat F}/\sqrt{m_b}$, 
which is the common value of $f_B$ 
and $f_{B^*}$ (modulo logarithmic factors) 
in the limit $m_b \to \infty$. The analysis is based on the
dominance of polar diagrams contributing to the process 
$B^- \to \mu^- {\bar \nu}_\mu \gamma$, the pole being
either the vector meson $B^*$ or  
the positive parity $J^P=1^+$ state  $B^\prime_1$
(see ref. \cite{noi} 
for further details).  
According to the analysis in \cite{noi}, 
in correspondence to 
the expected range of values of ${\hat F}$: 
${\hat F}\simeq 0.35 \; GeV^{3/2}$, the branching ratio
 ${\cal B}( B^- \to  \mu^- {\bar \nu}_\mu \gamma)$
should be ${\cal O}(10^{-6})$, which represents an 
enhancement with respect to the purely leptonic mode. 

In order to give further arguments in support of that analysis,
we want to 
consider the process (\ref{channel}) in a different context. 
More precisely,
whereas in \cite{noi} we have studied the feasibility of extracting $f_B={\hat 
F}/\sqrt{m_b}$ from future experimental data, in this letter we study 
the decay (\ref{channel}) within 
a well defined theoretical model in order to have an 
independent estimate of the decay rate.

We employ a relativistic 
constituent quark model already used to study several aspects of the 
$B$-meson phenomenology \cite{col,pietroni,col1}. 
Within this model 
the mesons are represented as bound states of valence quarks and antiquarks
interacting via a QCD inspired istantaneous potential 
with a linear dependence at large distances, to account for 
confinement, and a modified coulombic behaviour at short distances
to include the asymptotic freedom property of QCD. We adopt the
interpolating form between such asymptotic dependences provided by  
the Richardson potential \cite{richardson} 
\footnote{A smearing of the Richardson potential
at short distances has also been introduced  to take into account the effects 
of the relativistic kinematics; see
ref.\cite{pietroni} for the explicit form of the
potential.}.  In the rest frame, the state describing a $B_a$ meson
is represented as:
\be |B_a>=i \sum_{\alpha \beta} {\delta_{\alpha \beta} \over \sqrt{3}} 
\sum_{rs} {\delta_{rs} \over \sqrt{2}} 
\int d \vec{k}_1 \; \psi_B (\vec{k}_1) b^{\dag}(\vec{k}_1,r,\alpha)
d^{\dag}_a(-\vec{k}_1, s, \beta)|0> \hskip 3 pt, \label{eq : 5} 
\ee

\noindent where $\alpha$ and $\beta$ are colour indices, $r$ and $s$ are spin 
indices, $b^{\dag}$ and $d^{\dag}_a$ are creation operators of the quark $b$ 
and the antiquark ${\bar q_a}$, carrying momenta $\vec{k}_1$ and 
$-\vec{k}_1$ respectively.
The  $B$-meson wave function $\psi_B(\vec k_1)$ satisfies a wave equation with 
relativistic kinematics (Salpeter equation)\cite{salpeter} taking the form
(in the meson rest frame):
\be
\Big\{ \sqrt{\vec{k}_1^2 + m_b^2} +    
\sqrt{\vec{k}_1^2 + m_{q_a}^2}- M_{B_a} \Big\}
\psi_B( \vec{k}_1) 
+ \int d \vec {k^{\prime}_1} V(\vec{k}_1, \vec k_1^\prime) 
  \psi_B (\vec k_1^\prime)=0 \label{salp} \ee
\noindent
where $V(\vec{k}_1, \vec k_1^\prime) $ is the interaction potential in
the momentum space and
$\psi_B$ is covariantly normalized:
\be {1 \over (2\pi)^3} \int d \vec{k}_1 |\psi_B(\vec k_1)|^2=2 \; M_{B_a} 
\hskip 3 pt . \label{eq : 8} \ee

Solving eq.(\ref{salp}) by numerical methods one obtains the wave function; we 
shall present this result later on. It can be mentioned that by this model
a number of predictions have been 
derived; for example, the heavy meson spectrum, leptonic 
constants \cite{col,pietroni}, semileptonic
form factors and  strong decay constants \cite{col1}. 

In the framework of the relativistic QCD potential model 
the wave function $\psi_B$ 
is the main dynamical quantity governing the decay 
(\ref{channel}). As a matter of fact,
the amplitude of the process 
$B^-(p) \to \mu^-(p_1) \; {\bar \nu}_\mu(p_2) \; \gamma(k, \epsilon)$ 
can be written as
\be
{\cal A}(B^- \to \mu^- {\bar \nu}_\mu \gamma)
={G_F \over \sqrt{2}} V_{ub} \; \; \big( L^\mu \cdot \Pi_\mu \big) \hskip 3 pt 
, \label{amp} \ee 
\noindent where  
$G_F$ is the Fermi constant, $V_{ub}$ is the CKM matrix element involved in the 
decay, 
$L^\mu={\bar \mu}(p_1) \gamma^\mu (1-\gamma_5)\nu(p_2)$
is the weak leptonic current, and
$\Pi_\mu$ is defined by
$\Pi_\mu = \Pi_{\mu \nu} \epsilon^{* \nu}$
($\epsilon$ is the photon polarization vector);
$\Pi_{\mu \nu} $ represents the correlator
\be
\Pi_{\mu \nu}= i\; \int d^4 x 
e^{i q \cdot x} <0|T[ J_\mu(x) V_\nu(0) ]|B(p)>  \hskip 3 pt .  \label{pi} 
\ee
\noindent 
In eq.(\ref{pi})  $q$ is $q=p_1+p_2$,
$J_\mu(x)={\bar u}(x) \gamma_\mu (1 -\gamma_5) b(x)$
is the weak hadronic current and 
$V_\nu(0)={2 \over 3} e\; {\bar u}(0) \gamma_\nu u(0)-
{1 \over 3} e \; {\bar b}(0) \gamma_\nu b(0)$ is the electromagnetic (e.m.) 
current. The two pieces in the e.m.
current correspond to the 
coupling  to the light quark and to the heavy quark, respectively.
The corresponding contributions to $\Pi_{\mu \nu}$ will be 
referred to as $\Pi_{\mu \nu}^\ell$ and $\Pi_{\mu \nu}^{\it h}$,
depicted in figs. 1a,b.
$\Pi_{\mu \nu}^\ell$ contains the light quark propagator: 
\be 
S_u(x,0)=\int {d^4 \ell \over (2 \pi)^4} {e^{i \ell \cdot x} \over 
\ell^2-m_u^2} (\slash \ell +m_u) \label{prop} 
\ee
\noindent while $\Pi_{\mu \nu}^{\it h}$ contains the 
analogous $b$ quark propagator. 

The calculation of the time-ordered product appearing in (\ref{pi}) gives:
\be
\Pi_{\mu \nu}^\ell=-{2 \over 3} e\;  \int d^4 x \; 
e^{i q \cdot x} \int {d^4 \ell \over (2 \pi)^4} 
{ e^{i \ell \cdot x} \over \ell^2 - m_u^2} <0| \tilde J_{\mu \nu}^\ell|B> 
\hskip 3 pt . \label{ell} 
\ee
\noindent The analogous expression for $\Pi_\mu^{\it h}$ can be obtained by 
the replacements: $ {2 \over 3} \leftrightarrow -{1 \over 3}$, 
$m_u \leftrightarrow m_b$, 
$\tilde J_{\mu \nu}^\ell \leftrightarrow \tilde J_{\mu \nu}^{\it h}$.
The operators $\tilde J_{\mu \nu}^\ell $ and 
$\tilde J_{\mu \nu}^{\it h}$, which  depend on the integration variable $x$,
are written as
$\tilde J_{\mu \nu}^\ell = {\bar u}(0) \Gamma_{\mu \nu}^\ell b(x)$ 
and 
$\tilde J_{\mu \nu}^{\it h} = {\bar u}(x) \Gamma_{\mu \nu}^{\it h} b(0)$, 
with the $\Gamma$ matrices given by
$\Gamma_{\mu \nu}^\ell=\gamma_\nu 
(\slash \ell + m_u) \gamma_\mu (1 - \gamma_5)
$ and 
$\Gamma_{\mu \nu}^{\it h}= \gamma_\mu (1 - \gamma_5)
(\slash \ell + m_b) \gamma_\nu $. 
In the constituent quark model such operators 
can be expressed in terms of quark operators; for example one has:
\bea
\tilde J_{\mu \nu}^\ell &=&  \sum_{\alpha \beta} \sum_{r s} \delta_{\alpha \beta} 
\int { d^3 q_1 d^3 q_2 \over (2 \pi)^3} 
\Big[ { m_b m_u \over E_u({\vec q_1}) E_b({\vec q_2}) }\Big]^{1/2} 
\nonumber \\
&:& [ {\bar u}_u ({\vec q}_1,r) b_u^{\dag} ({\vec q}_1,r,\alpha) + 
{\bar v}_u({\vec q}_1,r) d_u({\vec q}_1,r,\alpha)] \; 
\Gamma_{\mu \nu}^\ell 
\nonumber \\
&& [u_b({\vec q}_2, s) b_b({\vec q}_2,s,\beta) e^{-i q_2 \cdot x} +
v_b({\vec q}_2, s) d_b^{\dag} ({\vec q}_2,s,\beta) e^{i q_2 \cdot x}]:
\label{corr}
\eea
\noindent where $E_q({\vec q})=\sqrt{{\vec q}^2+m_q^2}$ and $u_q$ ($v_q$) are 
quark (antiquark) spinors. 

By exploiting anticommutation relations among annihilation and creation 
operators, we obtain, in the $B$ meson rest frame:
\bea
&& \Pi_{\mu \nu}=
i {e \over \sqrt{6}}  \sum_r \int {d^3k_1 \over (2 \pi)^3} 
\psi_B({\vec k}_1) \Big[ { m_b m_u \over E_b({\vec k}_1) 
E_u({\vec k}_1) } \Big]^{1/2}  \label{ris} \\
&&{\bar v}_u (-{\vec k_1},r) 
\Bigg\{ -2\;  { \gamma_\nu [( \slash q_2 - \slash q) 
+m_u] \gamma_\mu (1 - \gamma_5) \over (q_2 - q)^2 -m_u^2} +
{\gamma_\mu (1 - \gamma_5)[( \slash q - \slash q_1) + m_b] \gamma_\nu \over 
(q_1-q)^2 - m_b^2} \Bigg\} u_b( {\vec k_1},r)  \nonumber
\eea
\noindent where  $q_1=(E_u, -{\vec k_1})$ and $q_2=(E_b, {\vec k_1})$. We 
recognize in the two factors in the curly brakets the contributions of 
$\Pi_{\mu\nu}^\ell$ and $\Pi_{\mu\nu}^{\it h}$, respectively.

Since 
$\Pi_{\mu \nu}$ only depends on two vectors, the $B$ meson momentum 
$p_\mu$ and the photon momentum $k_\mu$,  it can be written in terms of
six independent Lorentz structures:
\be 
\Pi_{\mu \nu}= \alpha \;  p_\mu p_\nu + \beta \;  k_\mu k_\nu + 
\zeta \; k_\mu p_\nu + 
\delta  \; p_\mu k_\nu + \xi \; g_{\mu \nu} + i \; \eta \; 
\epsilon_{\mu \nu \rho \sigma} p^\rho k^\sigma 
\hskip 3 pt . \label{decom} 
\ee
\noindent 
By  gauge invariance one has
$\alpha=0$ and $ \xi = - p \cdot k~ \zeta$; moreover, after saturation by 
 $\epsilon^{* \nu}$ one gets: 
\be
\Pi_\mu = \Pi_{\mu \nu} \epsilon^{* \nu} = [ \zeta\; 
( k_\mu p_\nu - p \cdot k g_{\mu \nu}) + i \; \eta
 \; \epsilon_{\mu \nu \rho \sigma} p^\rho k^\sigma ] \; \epsilon^{* \nu} 
\hskip 3 pt , \label{due}
\ee
\noindent i.e. only  the terms proportional to $\eta$ 
and $\zeta$ survive: they are 
the vector and the axial vector contribution, respectively.
It is convenient to compute eq.(\ref{due}) 
in the $B$ rest frame $p=(M_B, {\vec 0})$, with 
$k=(k^0,0,0,k^0)$; the result reads:
\be
\zeta={\Pi_{11} \over  M_B k^0} \hskip 1.5 cm 
\eta={ \Pi_{12} \over  i \;  M_B k^0} \hskip 3 pt .
\label{par} 
\ee
At this point, it is straightforward to calculate the rate of the decay process
(\ref{channel}); one obtains
($m_\mu \simeq 0$): 
\be 
\Gamma(B^- \to \mu^- {\bar \nu}_\mu \gamma)= {G_F^2 |V_{ub}|^2 \over 
3 (2 \pi)^3} \int_0^{M_B/2} dk^0 k^0 (M_B-2k^0)[|\Pi_{11}|^2 +
|\Pi_{12}|^2] \hskip 3 pt , \label{gamma} 
\ee
\noindent where:
\bea
\Pi_{11} &=& {i \;  e  \over 4 \sqrt{3} \pi } \int_{-1}^1 d cos \theta
\int_0^{|{\vec k}_1|_{max}} |{\vec k}_1| d |{\vec k}_1| 
\; u_B(|{\vec k}_1|) \Big[ {1 \over E_b 
E_u (E_b+m_b)(E_u+m_u)} \Big]^{1/2} \nonumber \\
\Bigg\{ &-& {2 \over f} \; \Big[ (M_B-k^0) [(E_b+m_b)(E_u+m_u)-|{\vec k}_1|^2 ]
+ |{\vec k}_1|^2 cos^2 \theta (E_u+m_u-E_b-m_b)  \nonumber \\
&+&|{\vec k}_1| k^0 cos \theta (E_u+m_u-E_b-m_b) \nonumber \\
&-& (E_b+m_b)(E_u+m_u)(E_b+m_u) + 
(E_b-m_u) |{\vec k}_1|^2 \Big] \label{1122} \\
&+& {1 \over g} \Big[ (M_B-k^0) [|{\vec k}_1|^2-(E_b+m_b)(E_u+m_u)] 
+ |{\vec k}_1|^2 cos^2 \theta (E_u+m_u-E_b-m_b)  \nonumber \\
&-&|{\vec k}_1| k^0 cos \theta (E_u+m_u-E_b-m_b) \nonumber \\
&+&(E_b+m_b)(E_u+m_u)(E_u+m_b) - 
(E_u-m_b) |{\vec k}_1|^2 \Big] \Bigg\} \nonumber
\eea
and
\bea
\Pi_{12}&=& {e \over 4  \sqrt{3} \pi } \int_{-1}^1 d cos \theta
\int_0^{|{\vec k}_1|_{max}} |{\vec k}_1| d |{\vec k}_1| 
\; u_B(|{\vec k}_1|) \Big[ {1 \over E_b 
E_u (E_b+m_b)(E_u+m_u)} \Big]^{1/2} \nonumber \\
\Big\{ && k^0 [(E_b+m_b)(E_u+m_u)- |{\vec k}_1|^2] + |{\vec k}_1| cos \theta 
(m_b-m_u)(E_u+m_u+E_b+m_b)  \nonumber  \\
&+& |{\vec k}_1| cos \theta (M_B-k^0)(E_u+m_u-E_b-m_b) \Big\} \Big( {2 \over f} 
- {1 \over g} \Big) \;\;\; .\label{1221}
\eea
\noindent 
In eqs.(\ref{1122},\ref{1221}) the quantities $f$ and $g$ are defined as
\be 
f=m_b^2+M_B^2 -2 M_B k^0 -2 M_B E_b +2 E_b k^0 -2 k^0 |{\vec k}_1| cos \theta 
-m_u^2 \label{f} 
\ee
\be 
g=m_u^2+M_B^2 -2 M_B k^0 -2 M_B E_u +2 E_u k^0 +2 k^0 |{\vec k}_1| cos \theta 
-m_b^2 \hskip 3 pt , \label{g} 
\ee
whereas the S-wave reduced  function $u$ is related to $\psi_B$:
\be
u_B(|{\vec k}_1|) = {|\vec k_1| \over \sqrt{2} \pi} \psi_B(\vec k_1) 
\;\; ; \label{u} 
\ee
we plot in fig.2 the function $u_B$ obtained as a solution of the wave equation
(\ref{salp}).

Let us now point out that in computing the diagrams in figs.1a,b and, 
therefore, in eqs.(\ref{1122}), (\ref{1221}), we have so far imposed 
4-momentum 
conservation for the physical particles $B$, $\mu$, $\nu$ and $\gamma$ 
in the process (\ref{channel}). On the other hand, energy conservation 
has to be imposed also at quark level since, otherwise, (\ref{1122}) and 
 (\ref{1221}) would present spurious kinematical singularities. 
In order to deal with this problem
we follow the approach originally proposed within the ACCMM model 
\cite{accmm} for the decay $b \to u \; \ell \; {\bar \nu}_\ell$. One assumes 
that the spectator
quark has a definite mass, while the active quark 
has a "running" mass, defined consistently with the energy conservation: 
\be
E_b+E_u=M_B \hskip 3 pt . \label{cons} 
\ee
Therefore, as in the case of the ACCMM model,
the  running mass of the active $b$ quark can be defined by: 
\be
m_b^2({\vec k_1})=M_B^2+m_u^2-2 M_B \sqrt{\vec k_1^2+m_u^2} \hskip 3 pt. 
\label{mb} \ee
\noindent 
Moreover,
by requiring that  the right hand side of eq. (\ref{mb}) 
is positive,  an upper bound on the quark momentum $|{\vec k}_1|$:
can be obtained 
\be
|{\vec k}_1| \le {M_B^2- m_u^2 \over 2 M_B} \; \; .\label{kmax} 
\ee
\noindent 
Notice that
the masses of the light constituent  quarks, as obtained by fits to the
meson spectrum, are $m_u=m_d=38 \hskip 3 pt MeV$.
\par
The contribution of the two physical processes when the photon couples to the 
light or to the heavy quark is still recognizable in eqs. 
(\ref{1122})-(\ref{1221}), 
since the quantities $f$ and $g$ come from the light and heavy 
quark propagators respectively. It turns out that the contribution of the terms 
proportional to ${1 \over g}$ are numerically much smaller than those 
proportional to ${1 \over f}$. This is not surprising, since the e.m. coupling 
of the photon to the quarks corresponds to a magnetic transition, 
and therefore it is inversely 
proportional to the quark mass.\par
A final remark concerns the  photon energy. 
In eq.(\ref{gamma}) we have allowed $k^0$ 
to vary in the range $[0,M_B/2]$; however the
integral diverges at the lower limit $k^0=0$.
This result is unphysical since it would correspond  to a zero energy
photon in the final state. Formally, this divergence would be canceled by 
radiative corrections to the formulae (\ref{1122}) and (\ref{1221}).
On the other hand, one should take into account 
that at future experiments, 
e.g. at the SLAC $B$-factory, the smallest measurable photon energy is 
of the order of $50 \; MeV$; therefore, it is a reasonable assumption to cut
off from the integral the small photon energies, and, at the same time, to 
neglect radiative corrections.

In our calculation, the effect of the unphysical divergence begins around a 
photon energy 
$k^0 \simeq 350 \; MeV$, and therefore we use this value as a lower bound 
for the photon energy. 

In fig.3 we plot the photon spectrum for the decay 
(\ref{channel}); 
the differential distribution has a peak around $1.5 \; 
GeV$, which should render it quite accessible to experimental analyses.

For the decay width we obtain the result 
 $\Gamma(B^- \to \mu^- {\bar \nu}_\mu \gamma)=3.7 \; 10^{-19} 
\big( {V_{ub} \over 3 \times 10^{-3}} \big)^2 \; GeV$, which 
corresponds to
\be
{\cal B}(B^- \to \mu^- {\bar \nu}_\mu \gamma) =0.9 \; 10^{-6}~~. \label{br} 
\ee

This result is obtained with a cut-off $\Delta=350 \; MeV$ in the 
photon energy. As it can be seen from fig.3, 
the uncertainty related to this choice should 
not be significant. For example, 
 putting $\Delta=100 \; MeV$ the result for the branching ratio increases by 
less than  $10 \%$. 

The conclusion we can draw from this result is
that the relativistic quark model gives predictions in agreement
with the expectations discussed in
ref. \cite{noi}; the weak radiative decay
$B^- \to \mu^- {\bar \nu}_\mu \gamma$
has an appreciable rate and might be observed in the near future. 

As for on the theoretical
uncertainties of the result (\ref{br}),
a part from the energy cut-off they mainly come from the choice of the $B$ 
meson wave function. A possible estimate of this
theoretical error consists in assuming 
a different wave function. A commonly used
quark momentum distribution inside the $B$ meson is given
by the ACCMM model \cite{accmm}; in our notations it corresponds 
to the gaussian wave function:
\be
u_B ( {\vec k}_1 ) = 2 |\vec k_1| 
\Big( {2 M_B \over \sqrt{\pi} P_F^3} \Big)^{1/2}
 exp \Big(  -{ {\vec k}_1^2 \over {2 P_F^2}}\Big ) \hskip 3 pt . \label{phi} 
\ee
\noindent The parameter $P_F$ is related to the heavy quark average square 
momentum: $<p^2>={3 \over 2} P_F^2$. In a recent analysis 
\cite{kim} this parameter has been fitted using
experimental data for
the inclusive decay $B \to X_c \ell {\bar \nu}_\ell$ 
\cite{cleo} with the result: $P_F=0.51 \; GeV$. 
Using (\ref{phi}) with such a value of $P_F$ in the previous
formulas, 
one would obtain ${\cal B}(B^- \to \mu^- 
{\bar \nu}_\mu \gamma)\simeq 0.8 \;\cdot 10^{-6}$.
This result suggests that the estimate given in
eq. ({\ref{br}) is rather accurate; 
the rate for the weak radiative decay $B^- \to \mu^- 
{\bar \nu}_\mu \gamma$ is large enough for a measurement 
at future accelerators.

\vskip 0.5cm
\par\noindent
Acknowledgements 
\par\noindent
We thank M.Carpinelli, M. Giorgi, A. Palano and N.Paver
for interesting  discussions.

\newpage
\begin{center} 
  \begin{Large}
  \begin{bf}
  Figure Captions
  \end{bf}
  \end{Large}
\end{center}
  \vspace{5mm}

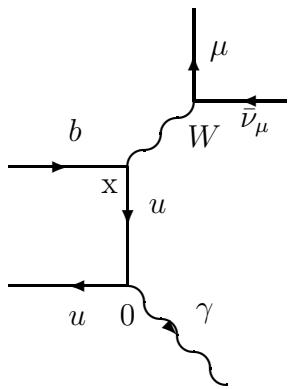
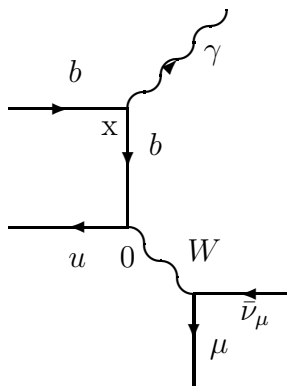
\noindent {\bf Figure 1}\\
\noindent
Diagrams describing the decay $B^- \to \mu^- {\bar \nu}_\mu \gamma$.

\vspace{5mm}
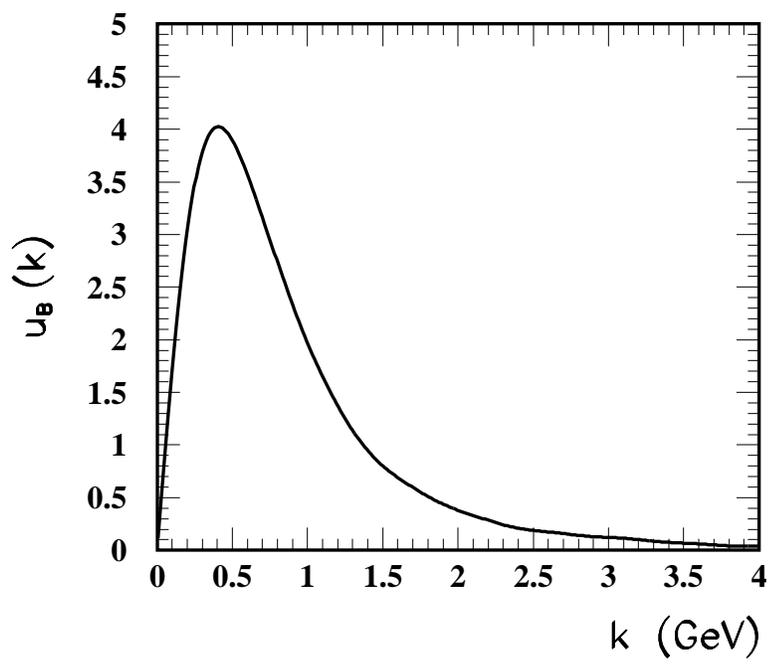
\noindent {\bf Figure 2}\\
\noindent
The wave function $u_B(k_1)$ as obtained by the QCD relativistic quark 
model.

\vspace{5mm}
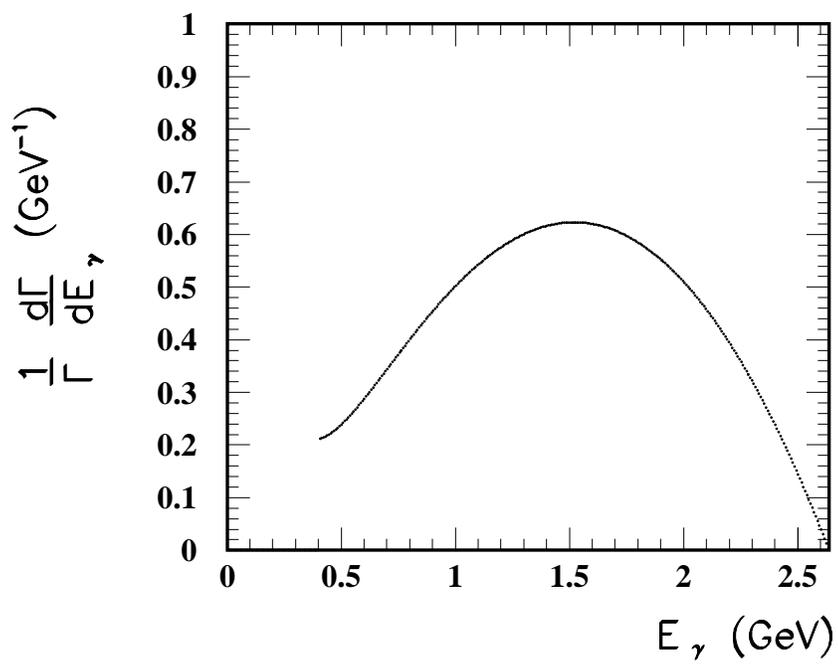
\noindent {\bf Figure 3}\\
\noindent
Predicted photon energy spectrum.

\vspace{5mm}

\newpage
\begin{center}
\input FEYNMAN
\begin{picture}(20000,15000)
\THICKLINES
\drawline\fermion[\E\REG](0,0)[4500]
\drawarrow[\W\ATTIP](\pmidx,\pmidy)
\global\advance\pmidy by -1500
\put(\pmidx,\pmidy){$u$}
\put(4200,-1500){0}
\drawline\photon[\SE\REG](\pbackx,\pbacky)[6]
\drawarrow[\LDIR\ATTIP](\pmidx,\pmidy)
\global\advance\pmidx by 700
\global\advance\pmidy by 700
\put(\pmidx,\pmidy){$\gamma$}
\drawline\fermion[\N\REG](\pfrontx,\pfronty)[4500]
\drawarrow[\S\ATTIP](\pmidx,\pmidy)
\global\advance\pmidx by 800
\global\advance\pmidy by 500
\put(\pmidx,\pmidy){$u$}
\put(3500,3500){x}
\drawline\fermion[\W\REG](\pbackx,\pbacky)[4500]
\drawarrow[\E\ATTIP](\pmidx,\pmidy)
\global\advance\pmidy by 1000
\put(\pmidx,\pmidy){$b$}
\drawline\photon[\NE\REG](\pfrontx,\pfronty)[4]
\global\advance\pmidx by 1000
\global\advance\pmidy by -500
\put(\pmidx,\pmidy){$W$}
\drawline\fermion[\N\REG](\pbackx,\pbacky)[3500]
\drawarrow[\LDIR\ATTIP](\pmidx,\pmidy)
\global\advance\pmidx by 600
\put(\pmidx,\pmidy){$\mu$}
\drawline\fermion[\E\REG](\pfrontx,\pfronty)[3500]
\drawarrow[\W\ATTIP](\pmidx,\pmidy)
\global\advance\pmidy by -1000
\put(\pmidx,\pmidy){${\bar \nu}_\mu$}
\end{picture}
\vskip 2 cm
{\bf Fig. 1 a}

\begin{picture}(20000,15000)
\THICKLINES
\drawline\fermion[\E\REG](0,0)[4500]
\drawarrow[\W\ATTIP](\pmidx,\pmidy)
\global\advance\pmidy by -1500
\put(\pmidx,\pmidy){$u$}
\put(4200,-1500){0}
\drawline\photon[\SE\REG](\pbackx,\pbacky)[4]
\global\advance\pmidx by 1000
\put(\pmidx,\pmidy){$W$}
\drawline\fermion[\S\REG](\pbackx,\pbacky)[3500]
\drawarrow[\LDIR\ATTIP](\pmidx,\pmidy)
\global\advance\pmidx by 600
\global\advance\pmidy by -500
\put(\pmidx,\pmidy){$\mu$}
\drawline\fermion[\E\REG](\pfrontx,\pfronty)[3500]
\drawarrow[\W\ATTIP](\pmidx,\pmidy)
\global\advance\pmidy by -1000
\put(\pmidx,\pmidy){${\bar \nu}_\mu$}
\drawline\fermion[\N\REG](\photonfrontx,\photonfronty)[4500]
\drawarrow[\S\ATTIP](\pmidx,\pmidy)
\global\advance\pmidx by 800
\global\advance\pmidy by 500
\put(\pmidx,\pmidy){$b$}
\put(3500,3500){x}
\drawline\fermion[\W\REG](\pbackx,\pbacky)[4500]
\drawarrow[\E\ATTIP](\pmidx,\pmidy)
\global\advance\pmidy by 1000
\put(\pmidx,\pmidy){$b$}
\drawline\photon[\NE\REG](\pfrontx,\pfronty)[6]
\drawarrow[\LDIR\ATTIP](\pmidx,\pmidy)
\global\advance\pmidx by 1000
\put(\pmidx,\pmidy){$\gamma$}
\end{picture}
\vskip 2.5 cm
{\bf Fig. 1 b}
\end{center}

\end{document}